\documentclass[11pt,twoside]{article}
\usepackage{IAUS212}
\usepackage{epsf}
\usepackage{rotating}

\def\edcomment#1{\iffalse\marginpar{\raggedright\sl#1\/}\else\relax\fi}
\reversemarginpar

\begin{document}
\vspace*{1cm}
\title{Inventory of Black Hole Binaries}
\author{Jerome A. Orosz}
\affil{Astronomical Institute, Utrecht University, The Netherlands}

\begin{abstract}
A small group of X-ray binaries currently provides the best evidence
for the existence of stellar-mass black holes.  These objects are
interacting binary systems where the X-rays arise from accretion of
material onto a compact object (i.e.\ an object with a radius of less
than a few hundred km).  In some favourable cases, optical studies of
the companion star lead to dynamical mass estimates for both
components.  In 17 cases, the mass of the compact object an X-ray
binary has been shown to exceed the maximum mass of a stable neutron
star (about $3\,M_{\odot}$), which leads to the conclusion that these
objects are black holes.  In this contribution I will review the basic
properties of these black hole binaries.
\end{abstract}

\section{Definition and Source Selection}

Black holes represent an extreme manifestation of Einstein's theory of
general relativity.  As an observational astronomer, I will not
consider any of the detailed theory of black holes but will instead
consider only the practical question of how to find them.  The usual
route is by the process of elimination.  In the current universe,
black holes must form via the gravitational collapse of a dying star.
Three outcomes are possible, depending on the mass of the degenerate
core: a white dwarf, a neutron star, or a black hole.  The mass of a
white dwarf cannot exceed the well-known Chandrasekhar limit, and a
neutron star has a somewhat analogous upper mass limit, generally
thought to be on the order of $3\,M_{\odot}$ (Rhoades \& Ruffini 1974;
Kalogera \& Baym 1996).  Once the mass of the degenerate core exceeds
about $3\,M_{\odot}$, no known force can halt the gravitational
collapse, and a black hole must be formed.  Given this, we have a
relatively straightforward observational definition of a black hole:
{\em A black hole is a compact object with a mass greater than three
solar masses}, where ``compact'' in this context means an object with
a radius smaller than about 100 km (i.e.\ much less than the radius of
normal stars).

Since black holes are dark, the only way one could hope to observe
them is through their gravitational influence on surrounding matter.
Early attempts to search catalogs of spectroscopic binaries to look
for single-lined binaries with massive and undetected companions were
not successful (Zel'dovich \& Guseynov 1966; Trimble \& Thorne 1969).
A far more efficient approach for source selection has its roots in
the 1960s when Zel'dovich and others realized that if a black hole
accreted material (either from a nearby companion star or the
interstellar medium), then it might shine brightly in X-rays and
$\gamma$ rays.  Today we know of a few hundred bright X-ray sources
which must be powered by accretion onto a compact object, either a
neutron star or a black hole (i.e.\ rapid variability in many cases
indicates a size scale of a few hundred km or less).  All of the black
hole candidates discussed below were selected on the basis of their
X-ray activity, although two of the sources, V404 Cyg and V4641 Sgr,
were previously known (optically) variable stars that were essentially
ignored until they were associated with bright X-ray sources.  Thus,
one should keep in mind that owing to special circumstances needed to
observe them (i.e.\ interacting binaries), the list of presently known
stellar-mass black holes is probably a very biased sample and may not
be representative of the general population of black holes.

\section{Mass Limits}

By my definition above, to establish the presence of a black hole in a
given system one must show that there is a compact object with a mass
greater than $3\,M_{\odot}$.  Establishing a lower limit to the mass
is relatively easy since in a single-lined spectroscopic binary, the
observed radial velocity curve of the visible star can be used to set
a firm lower limit on the mass of the unseen companion:
\begin{equation}
f(M)\equiv {PK_2^3\over 2\pi G}={M_1^3\sin ^3 i\over (M_1+M_2)^2},
\label{eq1}
\end{equation}
where $P$ is the binary orbital period, $K_2$ is the semiamplitude of
the companion's radial velocity curve, $G$ is the gravitation
constant, $i$ is the inclination of the orbit to the line of sight
($i=90^{\circ}$ for an orbit seen edge-on), and where $M_1$ and $M_2$
are the masses of the invisible object and the companion star,
respectively.  One can easily see that $f(M)$ (generally called the
``optical mass function'' or ``mass function'') is always {\em less
than} $M_1$ since $M_2>0$ and $\sin i\le 1$ always.  Thus a firm lower
limit on the mass of the compact object can be obtained from two
relatively simple measurements.

In real life, of course, things are always a bit more complicated.
For starters, in many cases one cannot even see spectral lines from
the companion star in many of the persistently X-ray bright sources.
Needless to say this precludes any attempts at dynamical mass
estimates.  In other cases, the companion star may not be uniformly
bright over its surface owing to heating by the X-ray source.  If so,
then the ``centre of light'' will not be the same as the ``centre of
mass'' and the observed radial velocity curve will be different than
the true one.  Since the value of the mass function depends on
$K_2^3$, small changes in $K_2$ can lead to relatively large changes
in $f(M)$.  Fortunately, many of the potential difficulties related to
X-ray heating are largely absent in the subset of the X-ray binaries
known as ``X-ray transients''.  As the name suggests, the X-ray
luminosity in these sources is highly variable (dynamic ranges of
$10^6$ to $10^7$ are not uncommon).  Dynamical observations of these
systems are generally reliable, provided one {\em observes the system
after the X-ray activity has completely ceased} (see Phillips,
Shahbaz, \& Podsiadlowski 1999; Shahbaz et al.\ 1999a; and Shahbaz et
al.\ 2000 for discussions regarding X-ray heating in GRO J1655-40).

Table 1 lists the mass functions for 17 systems that are widely
believed to contain black holes.  Fourteen of these systems are
transient X-ray sources with relatively low-mass companions ($M_2\la
1\,M_{\odot}$ generally), and the remaining three systems have massive
O/B companions whose bolometric luminosity is comparable to or exceeds
the X-ray luminosity of the accreting object.  Six of the systems have
mass functions which are above $\approx 5\,M_{\odot}$, and six more
have mass functions more than $\approx 3\,M_{\odot}$.  Based on the
mass functions alone, we can safely conclude these systems contain
black holes since the compact objects in them are too massive to be
neutron stars.  Although the other five systems have relatively small
mass functions, we can be reasonably sure these systems also contain
black holes since the currently best estimates for the compact object
masses (see the next section) are all above $\approx 3.5\,M_{\odot}$.
In addition to the sources listed in Table 1, there are a few dozen
sources which are believed to contain a black hole on the basis of
their X-ray spectra and temporal properties (e.g.\ Tanaka \& Lewin
1995; Liu, van Paradijs, \& van den Heuvel, 2001).

\section{Towards Actual Masses}

If possible, we want to do better than simply establishing a lower
limit to the mass of a black hole.  If we think of the companion star
as a test particle orbiting the black hole, then clearly in order to
measure the actual mass of the black hole after its mass function has
been measured, one must be able to convert {\em projected} radial
velocities into {\em actual} radial velocities (i.e.\ the inclination
$i$ is needed) and to find out how far the star is from the binary
centre-of-mass (i.e.\ one needs to know the mass ratio {\em or} the
mass of the companion star).

The procedure to derive astrophysically interesting parameters of a
binary system is conceptually simple.  One can make a model of a close
binary system that is specified by several parameters such as the
orbital period, the separation between the stars, the mass ratio, the
relative radii of the stars, the temperatures of the stars, the
inclination to the line of sight, etc.  Then, given this model, one
can compute various {\em observable} properties of this binary such as
the radial velocity curves, the light curves in various bandpasses,
eclipse duration(s) if any, the rotational velocities of the stars,
etc.  The goal then is to find a set of input model parameters that
can best match the set of observed properties of the binary of
interest.  Given the optimal input parameters, one can then compute
astrophysically interesting quantities such as the masses of the
components.

Models of close binary stars are relatively easy to construct, and
have been use for over thirty years (e.g.\ Wilson \& Devinney 1971;
Lyutyi, Syunyaev, \& Cherepashchuk 1973; Hutchings 1974; Avni \&
Bahcall 1975).  The difficult part has been related to the {\em
inverse} problem of parameter optimisation since the parameter space
to search is usually vast and often parameters can be tightly
correlated.  Common ways to tackle the inverse problem include
``differential corrections'' and related techniques (e.g.\ Wilson \&
Devinney 1971) and systematic ``grid searches'' (e.g.\ Kuiper, van
Paradijs, \& van der Klis 1988).  More recently, I have been using
optimisation techniques based on genetic algorithms (Orosz et al.\
2002a).  Genetic algorithms are very robust, and can allow for
accurate estimation of uncertainties on the fitted and derived
parameters.

In the case of the black hole binaries, the main observables usually
are a radial velocity curve for the companion star (this basically
sets the scale of the binary), a light curve for the companion in one
or more bandpasses (these mainly constrain the inclination), and in
most cases a measurement of the mean projected rotational velocity of
the companion star (this sets the mass ratio).  In nearly cases (with
the possible exception of SAX J1819.3-2525), the X-ray source is not
eclipsed, and one can set an upper limit to the inclination. Given
these observations, it is not an unduly difficult problem (in
principle!)  to find a good binary model for most of the black hole
systems listed in Table 1.

However, as before, in real life things are more complicated.  In many
cases, the observed light curve is not entirely due to the companion
star owing to the presence of extra sources of light (usually from an
accretion disk).  In the optical, the mean light curve can change from
one year to the next (e.g.\ Orosz et al.\ 1996; Leibowitz, Hemar, \&
Orio 1998; Pavlenko et al.\ 2001).  Presumably these long-term changes
are related to the accretion disk.  From spectroscopy, one can
estimate the light attributable to the accretion disk (e.g.\ Marsh,
Robinson, \& Wood 1994) and correct for it in the models.
Alternatively, it is generally {\em assumed} that the contamination
from the accretion disk is small in the $J$, $H$, and $K$ infrared
bands (e.g. Shahbaz et al.\ 1994; Gelino, Harrison, \& McNamara 2001).
There have a few attempts to directly measure the disk contamination
in the infrared with spectroscopy, but the results so far are not
terribly constraining (Shahbaz et al.\ 1996; Shahbaz, Bandyopadhyay,
\& Charles 1999b).

I have summarised in Table 1 the best (in my opinion) measurements of
the rotational velocities and inclinations for the 17 black hole
binaries.  Relatively uncertain values are denoted by '?'.  In some
cases there is reasonably good agreement between inclination
measurements made by different groups (e.g.\ GS 1124-683), while in
other cases there have been a wide range of measurements (e.g.\ Cyg
X-1, GRO J0422+32).  I have computed $\approx 1\sigma$ ranges for the
black hole mass and mass ratio using a simple Monte Carlo code that
makes use of numerical Roche lobe integrations to map a rotational
velocity into a mass ratio.  In some cases, rotational velocities are
not available and instead I have guessed the mass of the companion in
order to arrive at a mass range for the black hole.  Figure 1 shows
graphically the black hole mass ranges and, for comparison, masses for
neutron stars. All things considered, many of the masses are known
reasonably well, but one should always keep in mind potential
systematic errors such as those related to unstable light curves, etc.

\begin{sidewaystable}
\caption{Basic system parameters for the 14 transient and 3
persistent black hole binaries.}
\begin{tabular}{lccccccc}
\tableline
Source & Period  & $f(M)$ & $V_{\rm rot} \sin i$  
       & inclination  & $Q\equiv M_1/M_2$  range & BH mass range & refs.    \\
       & (days)  &  ($M_{\odot}$) & (km s$^{-1}$) & (deg)
 &  & ($M_{\odot}$)  & \\
\tableline
GRO J0422$+$32 & 0.2121600(2) & $1.19\pm 0.02$ & $90^{+22}_{-27}$ 
& $44\pm 2$     & $3.2-13.2$   & $3.66-4.97$  & 1-3 \\  
A0620$-$00     & 0.3230160(5) &  $2.72\pm 0.06$  & $83\pm 5$       
& $40.8\pm 3.0$ & $13.3-18.3$  & $8.70-12.86$  & 4-6 \\
GRS 1009$-$45  & 0.285206(2)  & $3.17\pm0.12$    & \dots           
& 67?           & $6.3-8.0$?   & $3.64-4.74$? & 7,8 \\			
XTE J1118$+$480& 0.169930(4)  & $6.1\pm 0.3$     & $114\pm 4$      
& $81\pm 2$      & $22.7-28.8$   &  $6.48-7.19$ & 9-11 \\ 		
GS 1124$-$683  & 0.432606(3)  & $3.01\pm 0.15$   & $106\pm 13$     
& $54\pm 2$      & $4.8-8.8$   & $6.47-8.18$  &  12-15 \\
4U 1543$-$47   & 1.116407(3)  & $0.25\pm 0.01$   & $46\pm 2$       
& $20.7\pm 1.5$   & $3.2-4.0$  & $8.45-10.39$ &     16 \\	   
XTE J1550$-$564& 1.5435(5)    & $6.86\pm 0.71$   & $90\pm 10$?     
& $72\pm 5$      & $>12$  & $8.36-10.76$ &      17 \\
GRO J1655$-$40 & 2.6219(2)    & $2.73\pm 0.09$   & $93\pm 3$       
& $70.2\pm 1.2$  & $2.4-2.7$  & $6.03-6.57$  &   18-20 \\
H1705$-$250    & 0.521(1)     & $4.86 \pm 0.13$  & $<79$           
& $>60$          & $>18.9$  & $5.64-8.30$ &   21-23 \\	
SAX J1819.3$-$2525& 2.81730(1)& $3.13\pm 0.13$   & $98.9\pm 1.5$   
& $75\pm 2$      & $2.22-2.39$  & $6.82-7.42$  &   24,25 \\		   
XTE J1859$+$226& 0.382(3)     & $7.4\pm 1.1$     & \dots           
& \dots         &  \dots  & $7.6-12.0$?  &      26 \\			   
GRS 1915$+$105 & 34(2)        & $9.5\pm 3.0$     & \dots           
& $70\pm 2$?     & \dots  & $10.0-18.0$? &    27-29\\			   
GS 2000$+$25   & 0.3440915(9) & $5.01\pm 0.12$   & $86\pm8$        
& $64.0\pm 1.3$  & $18.9-28.9$  & $7.15-7.78$ &   30-32 \\
GS 2023+338    & 6.4714(1)    & $6.08\pm 0.06$   & $38.8\pm 1.1$   
& $56\pm 4$      & $16.1-18.9$  & $10.06-13.38$  &   33,34 \\
                 &            &                  &    
&                &   &  &          \\
LMC X-3        & 1.70479(4)    & $2.29\pm 0.32$   & $130\pm 20$   
& $67\pm 3$      & $1.1-2.0$  & $5.94-9.17$ &   35-37 \\
LMC X-1        & 4.2288(6)    & $0.14\pm 0.05$   & \dots   
& $\approx 63$?      & $0.3-0.7$?  & $4.0-10.0$? &   38 \\
Cyg X-1    & 5.59983(2)    & $0.244\pm 0.005$   & $94\pm 5$   
& $35\pm 5$      & $0.50-0.57$  & $6.85-13.25$ &   39-42 \\
\tableline
\tableline
\end{tabular} \\
\spewtablenotes{{\bf References:} {\bf 1:}~Webb et al. 2000; 
{\bf 2:}~Harlaftis et al.\ 1999; 
{\bf 3:}~Gelino \& Harrison 2002;
{\bf 4:}~Leibowitz,  Hemar, \& Orio 1998;
{\bf 5:}~Marsh, Robinson, \& Wood 1994;
{\bf 6:}~Gelino, Harrison, \& Orosz 2001;
{\bf 7:}~Filippenko et al.\ 1999;
{\bf 8:}~Gelino  2002;
{\bf 9:}~Wagner et al.\ 2001;
{\bf 10:}~Orosz 2001;
{\bf 11:}~McClintock et al.\ 2001;
{\bf 12:}~Orosz et al.\ 1996;
{\bf 13:}~Casares et al.\ 1997;
{\bf 14:}~Shahbaz, Naylor, \& Charles 1997;
{\bf 15:}~Gelino, Harrison, \& McNamara 2001;
{\bf 16:}~Orosz et al.\ 2002b;
{\bf 17:}~Orosz et al.\ 2002a;
{\bf 18:}~Greene, Bailyn, \& Orosz 2001; 
{\bf 19:}~Shahbaz et al.\ 1999a;
{\bf 20:}~Israelian et al.\ 1999;
{\bf 21:}~Remillard et al.\ 1996;
{\bf 22:}~Filippenko et al.\ 1997;
{\bf 23:}~Harlaftis et al.\ 1997;
{\bf 24:}~Orosz et al.\ 2001;
{\bf 25:}~Orosz et al.\ 2002c;
{\bf 26:}~Filippenko \& Chornock 2001;
{\bf 27:}~Greiner,  Cuby, \& McCaughrean 2001;
{\bf 28:}~Mirabel \& Rodriguez 1994;
{\bf 29:}~Fender et al.\ 1999;
{\bf 30:}~Chevalier \& Ilovaisky 1993;
{\bf 31:}~Harlaftis, Horne, \& Filippenko 1996;
{\bf 32:}~Gelino 2001;
{\bf 33:}~Casares \& Charles 1994;
{\bf 34:}~Shahbaz et al.\ 1994;
{\bf 35:}~van der Klis et al.\ 1985;
{\bf 36:}~Cowley et al.\ 1983;
{\bf 37:}~Kuiper, van Paradijs, \& van der Klis 1988;
{\bf 38:}~Hutchings et al.\ 1987;
{\bf 39:}~Brocksopp et al.\ 1999;
{\bf 40:}~Ninkov, Walker, \& Yang 1987;
{\bf 41:}~Gies \& Bolton 1986;
{\bf 42:}~Herraro et al.\ 1995.
}
\end{sidewaystable}

\end{document}